\documentclass{ws-procs9x6}

\newcommand{\lwig}{\mbox{\ \raisebox{.3ex}
    {$<$}$\!\!\!\!\!$\raisebox{-.9ex}{$\sim$}\ }}
\newcommand{\gwig}{\mbox{\ \raisebox{.3ex}
    {$>$}$\!\!\!\!\!$\raisebox{-.9ex}{$\sim$}}\ }
\newcommand{\lambdabar}{{\hbox{$\lambda_e$\kern-1.9ex\raise+0.45ex\hbox{--}
\kern+0.2ex}}}

\begin{document}

\title{
\vspace{-3cm}
{\rm\normalsize\rightline{DESY 05-166}}
\vskip 1cm 
Extremely energetic cosmic neutrinos: 
Opportunities for astrophysics, particle physics, and cosmology 
\footnote{\uppercase{T}alk presented at
the \uppercase{ARENA} \uppercase{W}orkshop, \uppercase{DESY}, 
\uppercase{Z}euthen, \uppercase{G}ermany, \uppercase{M}ay 17-19, 2005.}
}

\author{Andreas Ringwald}

\address{Deutsches Elektronen-Synchrotron DESY,\\
Notkestra\ss e 85,\\ 
D-22607 Hamburg, Germany\\ 
E-mail: andreas.ringwald@desy.de}

\maketitle

\abstracts{
Existing and planned observatories for cosmic neutrinos open up a huge window in energy 
from $10^{7}$ to $10^{17}$~GeV.  Here, we discuss in particular   
the possibilities to use extremely energetic cosmic neutrinos as a diagnostic 
of astrophysical processes, as a tool for particle physics beyond the Standard Model, 
and as a probe of cosmology.
}

\section{Introduction}

\begin{figure}[t]
\begin{center}
\includegraphics*[bbllx=20pt,bblly=221pt,bburx=572pt,bbury=608pt,width=11cm]{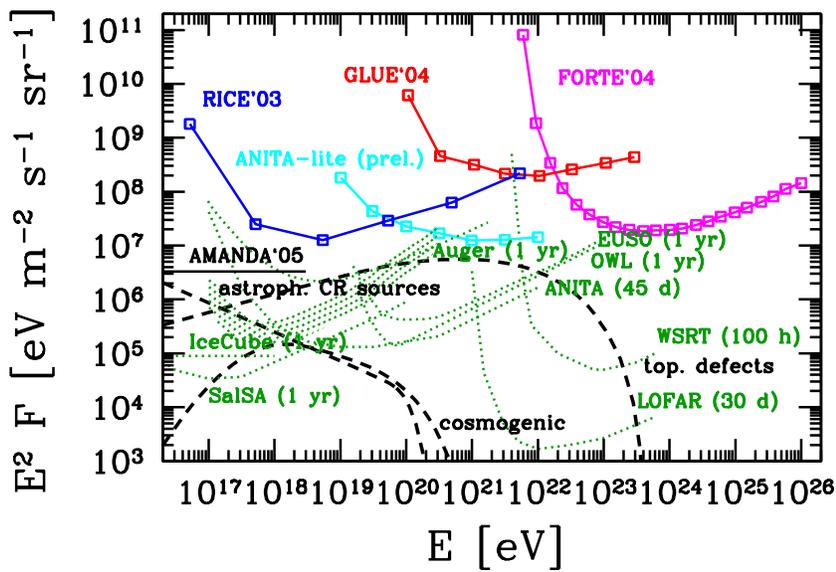}
\caption[...]{Current status and next decade prospects for EHEC$\nu$ physics, 
expressed in terms of diffuse neutrino fluxes per flavor, 
$F_{\nu_\alpha}+F_{\bar\nu_\alpha}$, $\alpha =e,\mu,\tau$.
Upper limits from  AMANDA\cite{Ackermann:2005sb}, ANITA-lite\cite{Barwick:2005arena}, 
FORTE\cite{Lehtinen:2003xv}, GLUE\cite{Gorham:2003da}, and RICE\cite{Kravchenko:2003tc}.
Also shown are projected sensitivities of 
ANITA\cite{Gorham:Anita}, EUSO\cite{Bottai:2003i}, IceCube\cite{Ahrens:2002dv}, 
LOFAR\cite{Scholten:2005pp}, OWL\cite{Stecker:2004wt}, 
the Pierre Auger Observatory in $\nu_e$, $\nu_\mu$ modes and in $\nu_\tau$ mode 
(bottom swath)\cite{Bertou:2001vm},  
SalSA\cite{Gorham:2001wr}, and WSRT\cite{Scholten:2005pp}, corresponding to 
one event per energy decade and indicated duration.   
Also shown are predictions from astrophysical CR sources\cite{Ahlers:2005sn}, 
from inelastic interactions of CR's with the cosmic microwave background (CMB) photons 
(cosmogenic neutrinos)\cite{Ahlers:2005sn,Fodor:2003ph}, 
and from topological defects\cite{fodor:tbp}.
\label{roadmap}} 
\end{center}
\end{figure}

We are living in exciting times for extremely high energy cosmic neutrinos 
(EHEC$\nu$'s). Existing observatories, such as AMANDA\cite{Ackermann:2005sb}, 
ANITA-lite\cite{Barwick:2005arena},   BAIKAL\cite{Wischnewski:2005rr}, 
FORTE\cite{Lehtinen:2003xv}, GLUE\cite{Gorham:2003da}, and RICE\cite{Kravchenko:2003tc} 
have recently put restrictive upper limits on the neutrino flux in the energy region
from $10^{7}$ to $10^{17}$~GeV (cf. Fig.~\ref{roadmap}). 
Furthermore, recent proposals for larger EHEC$\nu$ detectors, such as 
ANITA\cite{Gorham:Anita}, EUSO\cite{Bottai:2003i}, IceCube\cite{Ahrens:2002dv}, 
LOFAR\cite{Scholten:2005pp}, OWL\cite{Stecker:2004wt},  PAO\cite{Bertou:2001vm}, 
SalSA\cite{Gorham:2001wr}, WSRT\cite{Scholten:2005pp},
together with conservative neutrino flux predictions from astrophysical sources of the observed 
cosmic rays (CR's), such as active galactic nuclei, offer credible hope 
that the collection of a huge event sample above $10^{7}$~GeV may be realized within
this decade (cf. Fig.~\ref{roadmap}). 
This will provide not only important information on the astrophysical processes associated with the 
acceleration of CR's, but also an opportunity for particle physics beyond the reach of 
the Large Hadron Collider (LHC). 
There is even the possibility of a sizeable event sample above $10^{11}$~GeV, with important consequences 
for cosmology.  The corresponding neutrino fluxes may arise from the decomposition of topological defects  
--  relics of phase transitions in the very early universe -- into their particle constituents.  
Moreover, it may be possible to detect the  cosmic neutrino background 
via absorption features in these neutrino spectra. 
In this contribution, we will have a closer look at these exciting opportunities. 

\section{\boldmath EHEC$\nu$'s as a diagnostic of astrophysical processes}

\begin{figure}
\begin{center}
\parbox{5.cm}{\vspace{-6.8cm}\includegraphics[width=5.cm,clip=true]{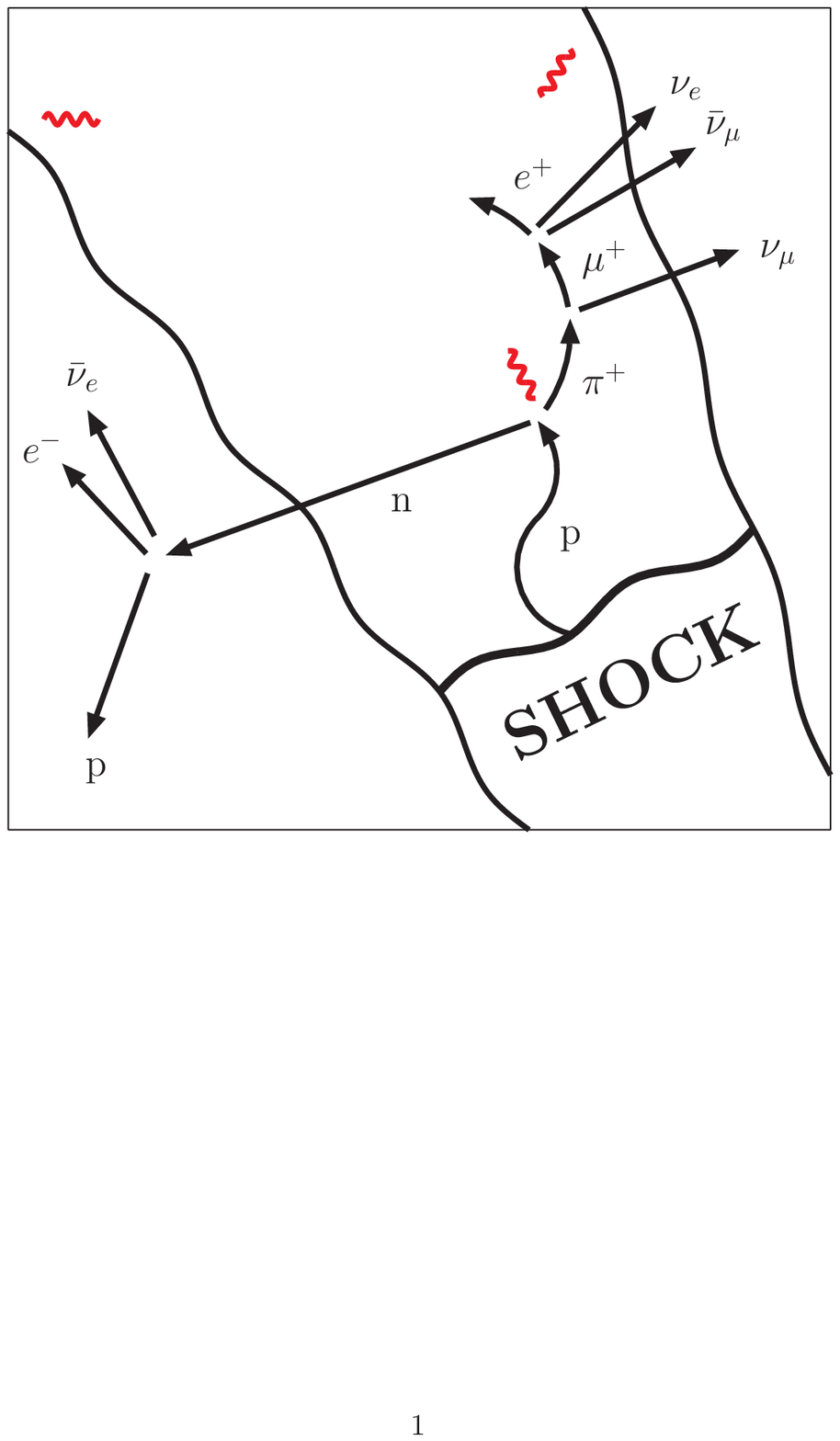}} 
\hfill
\includegraphics[width=6.cm]{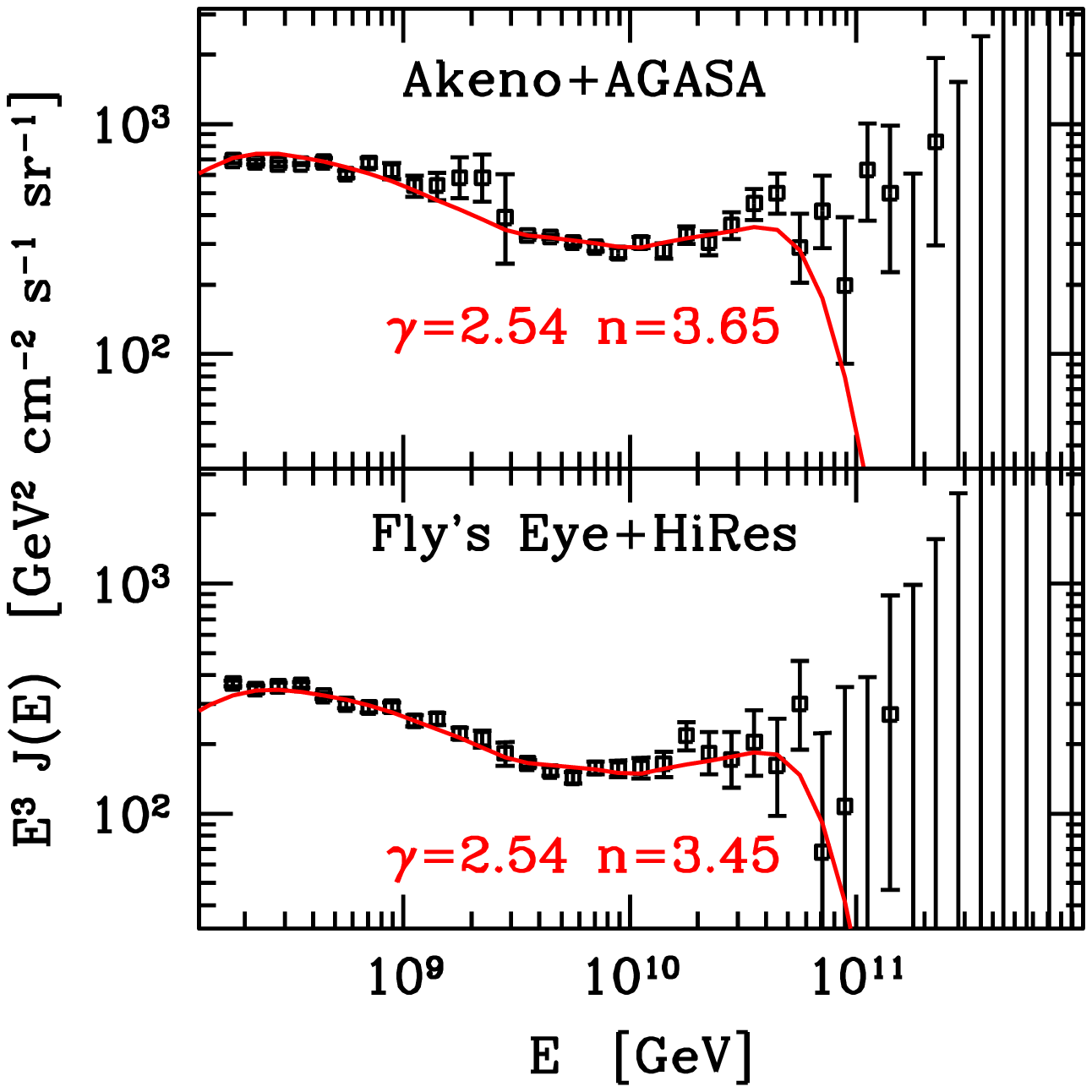}
\end{center}
\caption[]{\label{thin}
{\em Left:} Illustration of shock acceleration in the jet of an active
galaxy\cite{Ahlers:2005sn}.
{\em Right:} Best fits\cite{Ahlers:2005sn}
 to the ultra-high energy cosmic ray spectrum
in the energy interval $[10^{8.6},10^{11}]$~GeV as observed by 
Akeno\cite{Nagano:1991jz}+AGASA\cite{Takeda:1998ps} and Fly's
Eye\cite{Bird:1993yi}+HiRes\cite{Abu-Zayyad:2002sf}. 
The dip from $e^+e^-$ pair production\cite{Berezinsky:2002nc,Fodor:2003bn} 
and the bump from Greisen-Zatsepin-Kuzmin\cite{Greisen:1966jv} 
 (GZK) accumulation are clearly visible 
in the data and support the simple power law 
ansatz for the emissivity of the extragalactic sources, in which we have 
set $z_{\rm min} = 0.012$, $z_{\rm max} = 2$, and
$E_{i,{\rm max}} = 10^{12.5}$~GeV, for the boundaries in redshift and injection 
energy, respectively. Apparently, this fit undershoots the
data for the few highest energy events.}
\end{figure}

Neutrinos with energies $\lwig 10^{12}$~GeV propagate essentially without interaction between
their source and Earth. Hence, they are a powerful probe of high energy astrophysics, 
in particular of the conjectured acceleration sites of the CR's, notably active galactic nuclei (AGN).
A paradigm for the acceleration mechanism in the jets of these AGN's is shock acceleration.   
Protons and heavier nuclei are 
confined by magnetic fields and accelerated through repeated scattering 
by plasma shock fronts. Inelastic collisions of the trapped protons with the 
ambient plasma produces pions and neutrons, the former decaying into
neutrinos and photons, the latter eventually diffusing from the source and
decaying into CR protons (cf. Fig.~\ref{thin} (left)). 

A quite conservative estimate of the flux of neutrinos from such astrophysical sources 
can be made as follows\cite{Ahlers:2005sn}. Assuming that the sources are optically thin, i.e. the neutrons
can escape, one may determine the neutron emissivity at the sources from the 
observed CR spectra\cite{Waxman:1998yy}, taking into account propagation effects,
in particular $e^+e^-$ and pion production through inelastic scattering off the 
CMB photons.    
Figure~\ref{thin} (right) illustrates that both the AGASA and the 
HiRes data in the $10^{8.6\div 11}$~GeV range can be fitted nicely under the 
assumption of a simple power law neutron injection emissivity, $\propto E_i^{-2.5} (1+z)^{3.5}$,      
of the extragalactic sources, supporting the recent proposal towards a low transition energy, $\sim 10^{8.6}$~GeV, 
between galactic and extragalactic cosmic rays\cite{Berezinsky:2002nc}, which is also sustained by chemical composition
studies of HiRes data\cite{Bergman:2004bk}.  
The neutron injection emissivity is simply related to the 
neutrino emissivity, and the latter can be translated easily into an expected
neutrino flux at Earth. It should be detected very soon, 
if not already with AMANDA-II, then at least with IceCube (cf. Fig.~\ref{roadmap}),
which therefore can provide significant clues in demarcating the cosmic ray 
galactic/extragalactic crossover energy\cite{Ahlers:2005sn}. 
Although the cosmogenic neutrino flux from the inelastic interactions with the CMB photons 
starts to dominate over the neutrino flux from optically thin cosmic ray 
sources at energies above a few EeV, it appears to be hard to detect with the  
EHEC$\nu$ detectors operating in the next decade (cf. Fig.~\ref{roadmap}).  

\begin{figure}[t]
\begin{center}
\includegraphics*[bbllx=25pt,bblly=191pt,bburx=585pt,bbury=674pt,width=5.5cm]{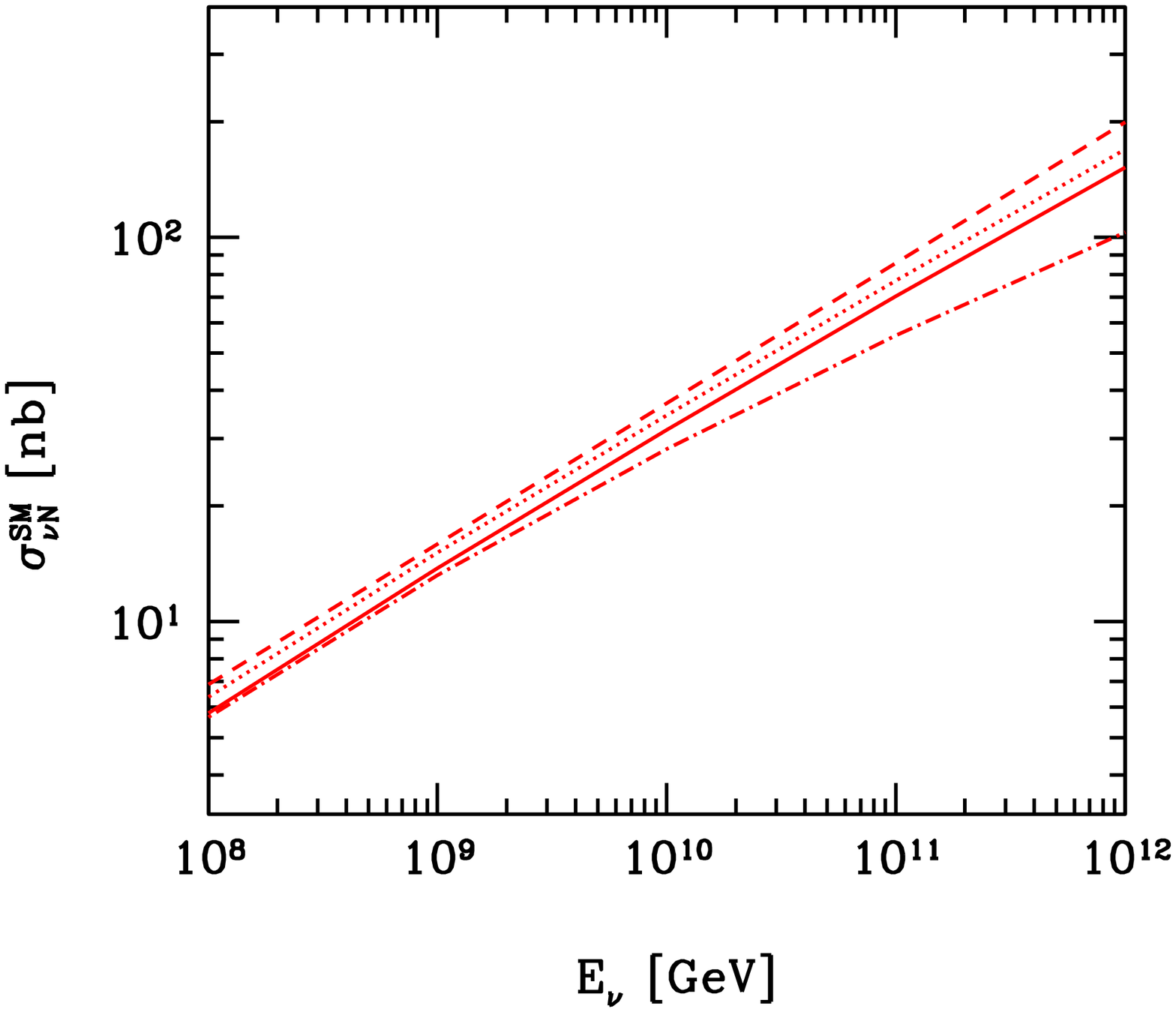}
\hfill
\parbox{5.8cm}{\vspace{-4.3cm}\includegraphics*[bbllx=22pt,bblly=172pt,bburx=566pt,bbury=578pt,width=5.8cm,clip=]{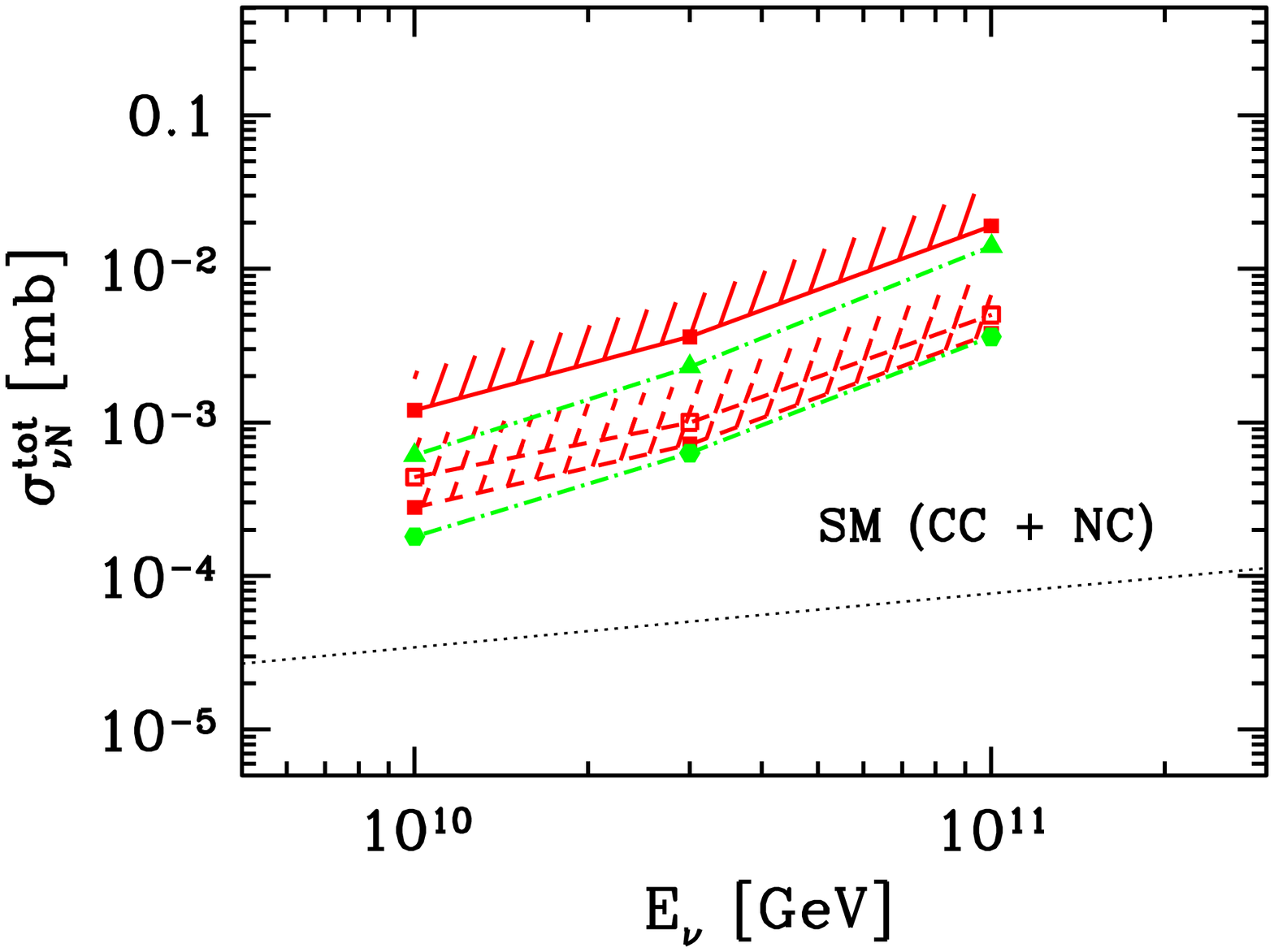}}
\caption[...]{
{\em Left:}
Standard Model $\nu N$ total cross section $\sigma_{\nu N}^{\rm tot}$ 
at extremely high neutrino energies $E_\nu$ obtained by various  
perturbative QCD resummation techniques\cite{Tu:2004ms}:
from a unified BFKL-DGLAP approach\cite{Kwiecinski:1998yf} (solid), 
based on CTEQ parton distributions\cite{Gandhi:1998ri} (dotted), 
based on GRV dynamical partons\cite{Gluck:1998js} (dashed), and 
from a unified BFKL-DGLAP approach supplemented by saturation effects\cite{Kutak:2003bd} (dashed-dotted).  
{\em Right:} 
Model-independent upper bounds on the neutrino-nucleon inelastic cross 
section\cite{Anchordoqui:2004ma}  derived from the
RICE Collaboration search results\cite{Kravchenko:2003tc}, by exploiting different 
cosmogenic neutrino flux estimates, by Fodor {\em et al.} (FKRT\cite{Fodor:2003ph}) (solid line) and
Protheroe and Johnson (PJ\cite{Protheroe:1996ft}) (dashed line joining solid squares).  
The dashed line joining the open squares (PJ) indicates the upper bound for 
inelasticity $\left< y \right> =0.5$. 
The dashed-dotted lines indicate the sensitivity (95\% CL, for  
$\sigma_{\nu N}^{\rm tot} <4$~mb) 
of PAO in 10 yr of operation assuming zero events observed above SM background
(circles PJ, triangles FKRT). 
For comparison, also shown is the SM total (charged current and 
neutral current) $\nu N$ inelastic cross 
section\cite{Gandhi:1998ri}. 
\label{SM}} 
\end{center}
\end{figure}

\section{\boldmath EHEC$\nu$'s and physics beyond the Standard Model}

Cosmic neutrinos with energies $E_\nu$ above $10^{8}$~GeV probe  
neutrino-nucleon scattering at center-of-mass (c.m.) energies above 
\begin{equation} 
\sqrt{s_{\nu N}}\equiv \sqrt{2m_NE_\nu} \simeq 14\ {\rm TeV}
\left( {E_\nu}/{10^{8}\ {\rm GeV}}\right)^{1/2}
\,, 
\end{equation}
beyond the proton-proton c.m. energy  $\sqrt{s_{pp}}=14$~TeV of the LHC, 
and Bjorken $x\equiv Q^2/(y\, s_{\nu N})$ values below
\begin{eqnarray}
x \simeq  2\times 10^{-4}\  
\left( {Q^2}/{m_W^2}\right)  
\left( {0.2}/{y}\right) 
\left( {10^{8}\ {\rm GeV}}/{E_\nu}\right)
\,,
\end{eqnarray}
where $Q^2$ is the momentum transfer squared, $m_W\simeq 80$~GeV the $W$-boson mass, 
and $y$ the inelasticity parameter. 
Under these kinematical conditions, the predictions for $\nu N$ scattering 
from the perturbative Standard Model (SM) are quite safely under control (cf. Fig.~\ref{SM} (left)), 
notably thanks to the input from measurements of deep-inelastic $ep$ scattering at 
HERA\cite{Adloff:2003uh,Chekanov:2003vw}. 
This makes it possible to search for enhancements in the $\nu N$ cross section 
due to physics beyond the (perturbative) SM, such as  
electroweak sphaleron\cite{Aoyama:1986ej} (non-perturbative $B+L$ violation),  
or Kaluza-Klein, black hole, $p$-brane, or string ball production in TeV scale gravity 
models\cite{Antoniadis:1990ew}.

Since the rate of neutrino-initiated showers is proportional to integrated flux
times cross section, 
the non-observation of quasi-horizontal or deeply-penetrating neutrino-induced air showers
as reported by, e.g., Fly's Eye\cite{Baltrusaitis:mt},  AGASA\cite{Yoshida:2001}, 
and RICE\cite{Kravchenko:2003tc} 
can be turned into an upper bound on the neutrino nucleon cross section if a certain 
prediction for the neutrino flux is 
exploited\cite{Berezinsky:kz,Morris:1993wg}.  
This is exemplified in Fig.~\ref{SM} (right), which displays 
the limits on $\sigma_{\nu N}$ from the RICE search results on contained showers\cite{Anchordoqui:2004ma},   
for two different assumptions about the EHEC$\nu$ flux. These bounds are considerably higher
than the SM cross section, albeit in the post-LHC energy region.   
PAO will be able to improve these limits by one order of magnitude\cite{Anchordoqui:2004ma}. 

\begin{figure}
\begin{center}
  \includegraphics[width=10.5cm,clip=true]{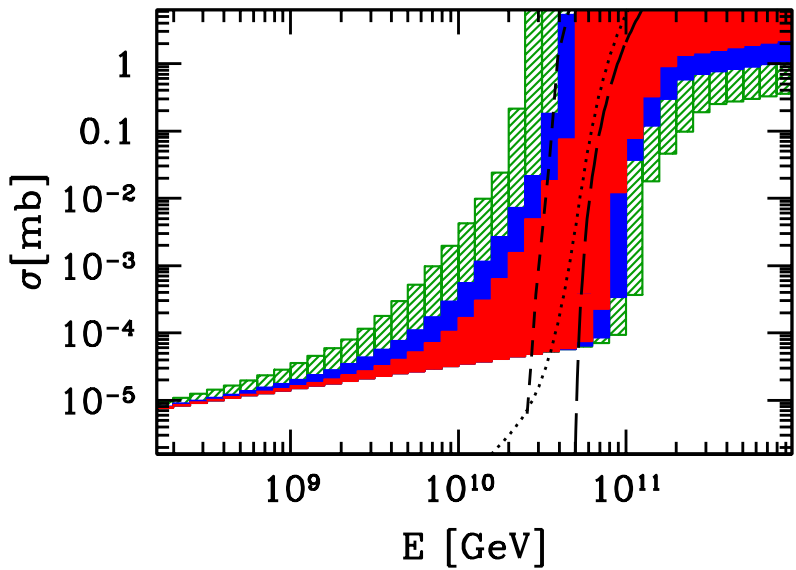}
\end{center}
\caption[]{\label{csbound} 
  The range of the cross section within the 99\%, 95\% and 90\% CL 
  required  
  for a successful strongly interacting neutrino scenario\cite{Ahlers:2005zy}. 
  The lines are theoretical predictions of an enhancement of the
  neutrino-nucleon cross-section by electroweak
  sphalerons\cite{Ringwald:2003ns,Han:2003ru} (short-dashed),
  $p$-branes\cite{Anchordoqui:2002it} (long-dashed) and string
  excitations\cite{Burgett:2004ac} (dotted).  }
\end{figure} 

The bounds exploiting searches for deeply-penetrating particles are 
typically applicable as long as $\sigma_{\nu N}\lwig 0.5\div 1$~mb. 
Models with even higher and more speculative cross sections,  
$\gwig 1\div10$~mb, such as  
electroweak sphaleron production, brane production, or 
string resonance production, qualify as  
strongly interacting neutrino scenarios\cite{Beresinsky:1969qj,Fodor:2003bn}, 
according to which the mysterious EHECR beyond the 
predicted GZK cutoff~\cite{Greisen:1966jv}  at 
$E_{\rm GZK}\simeq 4\times 10^{10}$~GeV (cf. Fig.~\ref{thin} (right))
are initiated by cosmogenic neutrinos. Figure~\ref{csbound} illustrates
that a combined fit of the existent data on vertical showers by 
AGASA and HiRes, as well as of the search results on weakly interacting particles 
of AGASA and RICE, requires a steep increase within one energy decade around 
$E_\mathrm{GZK}$ by four orders of magnitude\cite{Ahlers:2005zy} -- an enhancement 
which has indeed been proposed within some extensions of the (perturbative) SM.  

We have emphasized here the current constraints from EHEC$\nu$ on physics beyond the SM. 
A more detailed account of the particle physics reach of the planned EHEC$\nu$ observatories 
can be found elsewhere\cite{Han:2004kq,Anchordoqui:2005ey}. 

\section{\boldmath EHEC$\nu$'s as a tool to study big bang relics}

The existence of topological stable solutions of the field equations (topological defects) is a 
generic prediction of symmetry breaking (SB) in Grand Unified Theories (GUT's)
and occurs even at the fundamental level in String Theory in the form 
of F- and D-strings\cite{Polchinski:2004ia}. Specifically, $\rm G\to H\times U(1)$ SB leads to monopoles, 
$\rm U(1)$ SB to  ordinary or superconducting strings, and 
${\rm G\to H\times U(1)\to H\times Z}_N$ SB to  monopoles connected by strings, e.g.
necklaces in case of $N=2$. Such topological defects may be produced through 
non-thermal phase transitions during preheating after inflation\cite{Tkachev:1998dc}. 
Their superheavy constituents $X$, often gauge or Higgs bosons with 
masses $m_X\sim 10^{12\div 16}$~GeV, may be liberated on various occasions\cite{Bhattacharjee:1998qc}, 
e.g. through repeated self-intersections of strings, through annihilation of monopole antimonopole
pairs etc., and rapidly decay into stable SM particles, under which 
we readily find\cite{Bhattacharjee:1991zm} EHEC$\nu$'s with energies up to $\sim 0.05\,m_X$. 
The corresponding fragmentation spectra are meanwhile worked out very 
accurately\cite{Aloisio:2003xj} via Monte Carlo generators\cite{Birkel:1998nx} 
or via 
DGLAP evolution\cite{Fodor:2000za} 
 from  
experimentally determined initial distributions at the scale $m_Z$ to
the ones at $m_X$. The injection rate, which determines
in particular the overall normalization of the neutrino flux, depends on cosmic time $t$ in  
the form $\dot n_X = \kappa m_X^p t^{-4+p}$, where $\kappa$ and $p$ are 
dimensionless constants depending on the specific scenario\cite{Bhattacharjee:1991zm}.    

\begin{figure}
\vspace{-.5cm}
\begin{center}
  \includegraphics[bbllx=20pt,bblly=228pt,bburx=572pt,bbury=603pt,width=11.cm,clip=true]{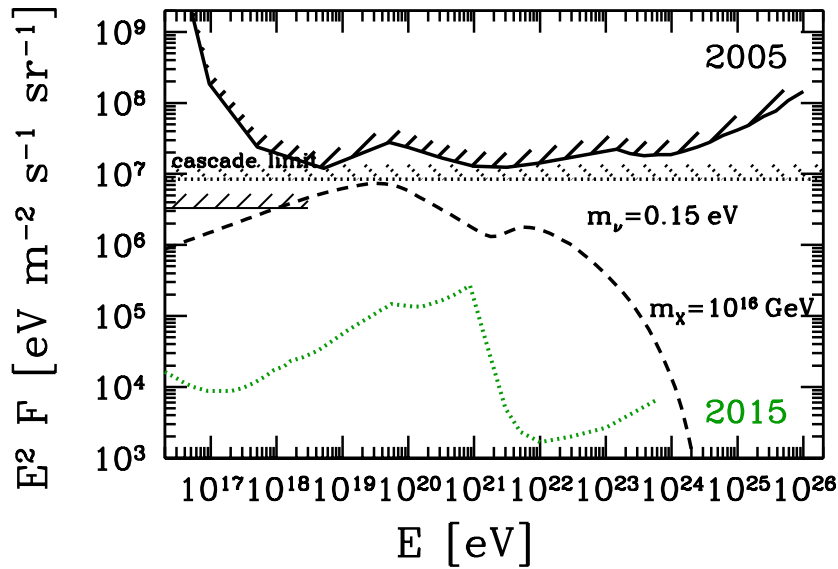}
\end{center}
\caption[]{\label{abs_dip} 
Present (2005) limits on the neutrino flux and projected sensitivity 
in ten years from now (2015), together with a prediction from 
topological defects\cite{fodor:tbp} ($m_X=10^{16}$~GeV, $p=0$). The absorption dip 
arising from resonant annihilation of the EHEC$\nu$'s with big
bang relic neutrinos of mass $m_\nu =0.15$~eV into $Z$-bosons 
is clearly visible.} 
\end{figure} 

For a wide range of overall flux normalizations, the upcoming EHEC$\nu$ observatories
seem to be sensitive enough to obtain, within the next decade, sizeable 
event rates from topological defects\cite{fodor:tbp} (cf. Fig.~\ref{abs_dip}). Note, that, for the   
first time in cosmic particle physics, the GUT energy scale can be directly probed.
Clearly, a precise measurement of the neutrino spectrum from topological defects
would have a strong impact on particle physics and cosmology. Its mere existence
would signal the existence of topological defects as 
relics from early phase transitions after inflation.  
The high end of the spectrum directly reveals the mass of the $X$ particles, and 
its shape entails detailed information on the particle content of the desert, 
on the Hubble expansion rate, and on the big bang relic neutrino background. 
Indeed, as illustrated in Fig.~\ref{abs_dip}, the resonant annihilation of 
the neutrinos from $X$ particle decays with big bang relic neutrinos would 
leave its imprints as absorption dips in the measured spectrum\cite{Weiler:1982qy}. 
Such a measurement would not only shed light on the existence and the 
spatial distribution\cite{Singh:2002de} 
of the cosmic neutrino background, 
but would also give important information on the neutrino masses\cite{Eberle:2004ua}, 
since the dips occur around the resonance energies 
$E_{\nu_i}^{\rm res}=4\times 10^{21}\ {\rm eV}\,(1\ {\rm eV}/m_{\nu_i})$.   
Note, that, along with a prediction of absorption dips, there goes a prediction of 
emission features\cite{Fargion:1997ft} 
-- protons and photons from hadronic $Z$-decay (``$Z$-bursts") -- 
which may appear as a CR flux recovery beyond $E_{\rm GZK}$ and be measured by 
EUSO, OWL, or LOFAR\cite{fodor:tbp}. 

\section{Conclusions}

The future seems bright in extremely high energetic neutrinos. 
There are many observatories under construction, whose combined sensitivity
ranges from $10^7$ to $10^{17}$~GeV, the energy scale of Grand Unification. 
In the likely case that appreciable event samples are collected
in this energy range, we can expect 
a strong impact on astrophysics, particle physics, and cosmology.


\begin{thebibliography}{0}

\bibitem{Ackermann:2005sb}
M.~Ackermann {\it et al.} [AMANDA Collab.],
Astropart.\ Phys.\  {\bf 22} (2005) 339.

\bibitem{Barwick:2005arena}
S.~Barwick {\em et al.} [ANITA Collaboration], these proceedings and to appear.   

\bibitem{Wischnewski:2005rr}
R.~Wischnewski {\it et al.}  [Baikal Collaboration],
arXiv:astro-ph/0507698.

\bibitem{Lehtinen:2003xv}
N.~G.~Lehtinen,  P.~W.~Gorham, A.~R.~Jacobson and R.~A.~Roussel-Dupre,
Phys.\ Rev.\ D {\bf 69} (2004) 013008.

\bibitem{Gorham:2003da}
P.~W.~Gorham, C.~L.~Hebert, K.~M.~Liewer, C.~J.~Naudet, D.~Saltzberg and D.~Williams,
Phys.\ Rev.\ Lett.\  {\bf 93} (2004) 041101.

\bibitem{Kravchenko:2003tc}
I.~Kravchenko,
arXiv:astro-ph/0306408.

\bibitem{Gorham:Anita}
P.~Gorham {\em et al.} [ANITA Collaboration], NASA Proposal  
SMEX03-0004-0019.

\bibitem{Bottai:2003i}
S.~Bottai and S.~Giurgola [EUSO Collaboration],
in: {\em Proc. 28th International Cosmic Ray Conference}, Tsukuba, Japan, 2003,
pp. 1113-1116; 
%
S.~Bottai [EUSO Collaboration],
to appear in: {\em Proc. Incontro Nazionale di Astrofisica delle Alte Energie}, 
Roma, 
2003.  

\bibitem{Ahrens:2002dv}
J.~Ahrens {\it et al.}  [IceCube Collab.],
Nucl.\ Phys.\ Proc.\ Suppl.\  {\bf 118} (2003) 388.

\bibitem{Scholten:2005pp}
O.~Scholten, J.~Bacelar, R.~Braun, A.~G.~de Bruyn, H.~Falcke, B.~Stappers and R.~G.~Strom,
arXiv:astro-ph/0508580.


\bibitem{Stecker:2004wt}
F.~W.~Stecker, J.~F.~Krizmanic, L.~M.~Barbier, E.~Loh, J.~W.~Mitchell, P.~Sokolsky and R.~E.~Streitmatter,
Nucl.\ Phys.\ Proc.\ Suppl.\  {\bf 136C} (2004) 433;
J.~F.~Krizmanic, private communications.

\bibitem{Bertou:2001vm}
X.~Bertou, P.~Billoir, O.~Deligny, C.~Lachaud and A.~Letessier-Selvon,
Astropart.\ Phys.\  {\bf 17} (2002) 183.

\bibitem{Gorham:2001wr}
P.~Gorham, D.~Saltzberg, A.~Odian, D.~Williams, D.~Besson, G.~Frichter and S.~Tantawi,
Nucl.\ Instrum.\ Meth.\ A {\bf 490} (2002) 476;
private commun. 

\bibitem{Ahlers:2005sn}
M.~Ahlers, L.~A.~Anchordoqui, H.~Goldberg, F.~Halzen, A.~Ringwald and T.~J.~Weiler,
Phys.\ Rev.\ D {\bf 72} (2005) 023001.

\bibitem{Fodor:2003ph}
Z.~Fodor, 
S.~D.~Katz, A.~Ringwald and H.~Tu,
JCAP {\bf 0311} (2003) 015.

\bibitem{fodor:tbp}
Z.~Fodor, S.~D.~Katz, A.~Ringwald, T.~J.~Weiler and Y.~Y.~Y.~Wong, 
DESY 05-165.


\bibitem{Waxman:1998yy}
E.~Waxman and J.~N.~Bahcall,
Phys.\ Rev.\ D {\bf 59} (1999) 023002;
%
K.~Mannheim, R.~J.~Protheroe and J.~P.~Rachen,
Phys.\ Rev.\ D {\bf 63} (2001) 023003.

\bibitem{Nagano:1991jz}
M.~Nagano {\it et al.},
J.\ Phys.\ G {\bf 18} (1992) 423.


\bibitem{Takeda:1998ps}
M.~Takeda {\it et al.} [AGASA Collab.],
Phys.\ Rev.\ Lett.\  {\bf 81}  (1998) 1163;
Astropart.\ Phys.\  {\bf 19}  (2003) 447;
http://www-akeno.icrr.u-tokyo.ac.jp/AGASA/

\bibitem{Bird:1993yi}
D.~J.~Bird {\it et al.}  [Fly's Eye Collaboration],
Phys.\ Rev.\ Lett.\  {\bf 71} 3401 (1993);
%
Astrophys.\ J.\  {\bf 424}, 491 (1994);
%
Astrophys.\ J.\  {\bf 441}, 144 (1995).

\bibitem{Abu-Zayyad:2002sf}
T.~Abu-Zayyad {\it et al.}  [HiRes Collaboration],
Astropart.\ Phys.\  {\bf 23} (2005) 157.

\bibitem{Berezinsky:2002nc}
V.~Berezinsky, A.~Z.~Gazizov and S.~I.~Grigorieva,
arXiv:hep-ph/0204357;
Phys.\ Lett.\ B {\bf 612} (2005) 147.


\bibitem{Fodor:2003bn}
Z.~Fodor, S.~D.~Katz, A.~Ringwald and H.~Tu,
Phys.\ Lett.\ B {\bf 561} (2003) 191. 



\bibitem{Greisen:1966jv}
K.~Greisen,
Phys.\ Rev.\ Lett.\  {\bf 16} (1966) 748;
%
G.~T.~Zatsepin and V.~A.~Kuzmin,
JETP Lett.\  {\bf 4} (1966) 78.

\bibitem{Bergman:2004bk}
D.~R.~Bergman  [HiRes Collab.],
Nucl.\ Phys.\ Proc.\ Suppl.\  {\bf 136} (2004) 40.

\bibitem{Tu:2004ms}
H.~Tu,
DESY-THESIS-2004-018;
unpubl.


\bibitem{Kwiecinski:1998yf}
J.~Kwiecinski,  A.~D.~Martin and A.~M.~Stasto,
Phys.\ Rev.\ D {\bf 59} (1999) 093002.

\bibitem{Gandhi:1998ri}
R.~Gandhi, C.~Quigg, M.~H.~Reno and I.~Sarcevic,
Phys.\ Rev.\ D {\bf 58} (1998) 093009.


\bibitem{Gluck:1998js}
M.~Gl\"uck, S.~Kretzer and E.~Reya,
Astropart.\ Phys.\  {\bf 11} (1999) 327.

\bibitem{Kutak:2003bd}
K.~Kutak and J.~Kwiecinski,
Eur.\ Phys.\ J.\ C {\bf 29} (2003) 521.

\bibitem{Adloff:2003uh}
C.~Adloff {\it et al.}  [H1 Collaboration],
Eur.\ Phys.\ J.\ C {\bf 30} (2003) 1.

\bibitem{Chekanov:2003vw}
S.~Chekanov {\it et al.}  [ZEUS Collaboration],
Eur.\ Phys.\ J.\ C {\bf 32} (2003) 1.

\bibitem{Aoyama:1986ej}
H.~Aoyama and H.~Goldberg,
Phys.\ Lett.\ B {\bf 188} (1987) 506;
%
A.~Ringwald,
Nucl.\ Phys.\ B {\bf 330} (1990) 1;
%
O.~Espinosa,
Nucl.\ Phys.\ B {\bf 343} (1990) 310.


\bibitem{Antoniadis:1990ew}
I.~Antoniadis,
{ Phys.\ Lett.}\ B {\bf 246} (1990) 377;
%
J.~D.~Lykken,
{ Phys.\ Rev.}\ D {\bf 54} (1996) 3693; 
%
N.~Arkani-Hamed, S.~Dimopoulos and G.~R.~Dvali,
{ Phys.\ Lett.}\ B {\bf 429} (1998) 263;
%
L.~Randall and R.~Sundrum,
{ Phys.\ Rev.\ Lett.}\  {\bf 83} (1999) 3370. 



\bibitem{Baltrusaitis:mt}
R.~M.~Baltrusaitis {\it et al.},
Phys.\ Rev.\ D {\bf 31} (1985) 2192.

\bibitem{Yoshida:2001}
S.~Yoshida {\em et al.} [AGASA Collaboration], 
in: {\em Proc. 27th International Cosmic Ray Conference}, Hamburg, Germany, 2001,
p. 1142

\bibitem{Berezinsky:kz}
V.~S.~Berezinsky and A.~Y.~Smirnov,
Phys.\ Lett.\ B {\bf 48} (1974) 269.


\bibitem{Morris:1993wg}
D.~A.~Morris and A.~Ringwald,
Astropart.\ Phys.\  {\bf 2} (1994) 43.

\bibitem{Protheroe:1996ft}
R.~J.~Protheroe and P.~A.~Johnson,
Astropart.\ Phys.\  {\bf 4} (1996) 253.

\bibitem{Anchordoqui:2004ma}
L.~A.~Anchordoqui, Z.~Fodor, S.~D.~Katz, A.~Ringwald and H.~Tu,
JCAP {\bf 0506} (2005) 013.

\bibitem{Beresinsky:1969qj}
V.~S.~Berezinsky and G.~T.~Zatsepin,
Phys.\ Lett.\ B {\bf 28} (1969) 423. 


\bibitem{Ahlers:2005zy}
M.~Ahlers, A.~Ringwald and H.~Tu,
arXiv:astro-ph/0506698.

\bibitem{Ringwald:2003ns}
A.~Ringwald,
JHEP {\bf 0310} (2003) 008.

\bibitem{Han:2003ru}
T.~Han and D.~Hooper,
Phys.\ Lett.\ B {\bf 582} (2004) 21.

\bibitem{Anchordoqui:2002it}
L.~A.~Anchordoqui,  J.~L.~Feng and H.~Goldberg,
Phys.\ Lett.\ B {\bf 535} (2002) 302.

\bibitem{Burgett:2004ac}
W.~S.~Burgett,  G.~Domokos and S.~Kovesi-Domokos,
Nucl.\ Phys.\ Proc.\ Suppl.\  {\bf 136} (2004) 327.

\bibitem{Han:2004kq}
T.~Han and D.~Hooper,
New J.\ Phys.\  {\bf 6} (2004) 150.

\bibitem{Anchordoqui:2005ey}
L.~Anchordoqui, T.~Han, D.~Hooper and S.~Sarkar,
arXiv:hep-ph/0508312.

\bibitem{Polchinski:2004ia}
J.~Polchinski,
arXiv:hep-th/0412244.


\bibitem{Tkachev:1998dc}
I.~Tkachev,  S.~Khlebnikov, L.~Kofman and A.~D.~Linde,
Phys.\ Lett.\ B {\bf 440} (1998) 262.

\bibitem{Bhattacharjee:1998qc}
P.~Bhattacharjee and G.~Sigl,
Phys.\ Rept.\  {\bf 327} (2000) 109.

\bibitem{Bhattacharjee:1991zm}
P.~Bhattacharjee,  C.~T.~Hill and D.~N.~Schramm,
Phys.\ Rev.\ Lett.\  {\bf 69} (1992) 567.

\bibitem{Aloisio:2003xj}
R.~Aloisio, V.~Berezinsky and M.~Kachelriess,
Phys.\ Rev.\ D {\bf 69} (2004) 094023.


\bibitem{Birkel:1998nx}
M.~Birkel and S.~Sarkar,
Astropart.\ Phys.\  {\bf 9} (1998) 297;
%
V.~Berezinsky and M.~Kachelriess,
Phys.\ Rev.\ D {\bf 63} (2001) 034007.

\bibitem{Fodor:2000za}
Z.~Fodor and S.~D.~Katz,
Phys.\ Rev.\ Lett.\  {\bf 86} (2001) 3224;
%
S.~Sarkar and R.~Toldra,
Nucl.\ Phys.\ B {\bf 621} (2002) 495; 
%
C.~Barbot and M.~Drees,
Astropart.\ Phys.\  {\bf 20} (2003) 5.


\bibitem{Weiler:1982qy}
T.~J.~Weiler,
Phys.\ Rev.\ Lett.\  {\bf 49} (1982) 234. 

\bibitem{Singh:2002de}
S.~Singh and C.~P.~Ma,
Phys.\ Rev.\ D {\bf 67} (2003) 023506;
%
A.~Ringwald and Y.~Y.~Y.~Wong,
JCAP {\bf 0412} (2004) 005.

\bibitem{Eberle:2004ua}
B.~Eberle,  A.~Ringwald, L.~Song and T.~J.~Weiler,
Phys.\ Rev.\ D {\bf 70} (2004) 023007; 
%
G.~Barenboim,  O.~Mena Requejo and C.~Quigg,
Phys.\ Rev.\ D {\bf 71} (2005) 083002; 
%
J.~C.~D'Olivo,  L.~Nellen, S.~Sahu and V.~Van Elewyck,
arXiv:astro-ph/0507333.

\bibitem{Fargion:1997ft}
D.~Fargion, B.~Mele and A.~Salis,
Astrophys.\ J.\  {\bf 517} (1999) 725;
%
T.~J.~Weiler,
Astropart.\ Phys.\  {\bf 11} (1999) 303;
%
Z.~Fodor, S.~D.~Katz and A.~Ringwald,
Phys.\ Rev.\ Lett.\  {\bf 88} (2002) 171101;
%
JHEP {\bf 0206} (2002) 046;
%
A.~Ringwald, T.~J.~Weiler and Y.~Y.~Y.~Wong,
Phys.\ Rev.\ D {\bf 72} (2005) 043008.

\end{thebibliography}
\end{document}